\documentclass[12pt,a4paper]{article}
\begin{document}
\begin{center}
{\Large \bf Squeezed and entangled collinear gluon states\\ at the non-perturbative stage of QCD jet evolution}\\
\vspace{4 mm}

            V.I.~Kuvshinov, V.A.~Shaparau

\vspace{4 mm}

Institute of Physics, National Academy
of Sciences of Belarus\\ 
F. Skaryna av. 68, 220072 Minsk, Belarus\\
\small\textit{E-mail: kuvshino@dragon.bas-net.by; shaporov@dragon.bas-net.by}\\

\end{center}

\begin{abstract}
At the non-perturbative stage of jet evolution, fluctuations of collinear gluons are less than those for coherent states that is indication of gluon squeezed states. We show that gluon entangled states which are closely related with two-mode squeezed states of gluon fields can appear by analogy with corresponding photon states in quantum optics.
\end{abstract}

\section{Introduction}
\hspace{10pt} Analogies between multiple hadron in HEP
and photon production in quantum optics (QO) discussed long ago \cite{sv_1}-\cite{Carruthers}. It was shown, in particular, that the general distribution that characterizes $e^+e^-, p\bar p, $ neutrino-induced collisions is a $k$-mode squeezed state distribution \cite{MD_SS,MD_SS_1}, the  multiplicity distribution of the pions has been explained by formalism of the squeezed isospin states \cite{S_isospin_S}.

Squeezed states (SS), introduced by Stoler \cite{sv_stoler}
and named by Hollenhorst \cite{sv_hol}, provoke great interest in connection with their uncommon properties: they can have
both sub-Poissonian (for coincide phases) and super-Poissonian (for antiphases) statistics corresponding
to antibunching and bunching of photons~\cite{Kilin}-\cite{QO_Cambr} and characteristic behaviour of the factorial and cumulant moments~\cite{PSS}. Moreover oscillatory behaviour of the multiplicity distribution of the photon SS \cite{QO_Japan} is different from Poissonian and Negative binomial distributions (NBD). The squeezed light is generated from the coherent one by nonlinear devices and is pure quantum non-perturbative phenomenon \cite{QO_Japan}-\cite{QO_Cambr}.

Studying non-perturbative evolution of gluon states prepared by perturbative
cascade stage in jets \cite{Soft} we have proved within Quantum chromodynamics (QCD) that this stage of jet evolution can be a source of gluon SS by analogy with nonlinear devices in QO for photon \cite{acta}-\cite{shaporov}. In these papers we investigated single-mode squeezing effect for virtual gluons with defined colour and vector components during small temporal evolution. Using the Local parton hadron duality it is easy to show that in this case behaviour of hadron multiplicity distribution in jet events is differentiated from the negative binomial one that is confirmed by experiments for $pp, p\overline{p}$-collisions~\cite{UA5}-\cite{OPAL}.

At the same time two-mode photon SS \cite{QO_Walls} in the limit of infinite squeezing are isomorphic to the Bell states \cite{Bell} which have been introduced in relation to the Einstein-Podolsky-Rosen (EPR) paradox \cite{EPR} and they are one of the examples of the entangled states for two polarizations \cite{Wolf}. The states
\begin{equation}
\label{Bell}
\begin{array}{l}
|\Phi^+\rangle=(|\!\updownarrow\rangle_{_{\!1}}|\!\updownarrow\rangle_{_{\!2}}
+|\!\leftrightarrow\rangle_{_{\!1}}|\!\leftrightarrow\rangle_{_{\!2}})/
\sqrt{2}, \qquad
|\Phi^-\rangle=(|\!\updownarrow\rangle_{_{\!1}}|\!\updownarrow\rangle_{_{\!2}}
-|\!\leftrightarrow\rangle_{_{\!1}}|\!\leftrightarrow\rangle_{_{\!2}})/\sqrt{2},\\
|\Psi^+\rangle=(|\!\updownarrow\rangle_{_{\!1}}|\!\leftrightarrow\rangle_{_{\!2}}
+|\!\leftrightarrow\rangle_{_{\!1}}|\!\updownarrow\rangle_{_{\!2}})/\sqrt{2}, \qquad
|\Psi^-\rangle=(|\!\updownarrow\rangle_{_{\!1}}|\!\leftrightarrow\rangle_{_{\!2}}
-|\!\leftrightarrow\rangle_{_{\!1}}|\!\updownarrow\rangle_{_{\!2}})/\sqrt{2}
\end{array}
\end{equation}
are the basis of the Bell states. Each of these entangled states, for example, $|\Phi^{\pm}\rangle,$ has a uncommon property: if one photon is registered with defined polarization (for example, with polarization $\updownarrow$), the other photon immediately becomes opposite polarized (longitudinal polarization). Thus a measurement over one particle has an instantaneous effect on the other, possibly located at a large distance. In particular, recently it was shown that it is possible for two photons to travel a total of 600 metres through free space and still remain "entangled" \cite{Austria}.

At finite squeeze factor  $r$ ($r$ is positive) a continuous variables entangled state is known from quantum optics as a two-mode squeezed state \cite{QO_Walls,Wolf}
\begin{equation}
\label{sq_2_mode}
  |{\rm f}\rangle = S(r) |0\rangle_1|0\rangle_2 = \frac{1}{\cosh r}\sum_{n=0}^\infty(\tanh
r)^{\! n}\,|n\rangle_1|n\rangle_2,
\end{equation}
where $S(r)=\exp\{r(a_1^+a_2^+ - a_1 a_2)\}$ is the operator of two-mode squeezing.
Rewriting the expression (\ref{sq_2_mode}) at small squeezing parameter $r$
\begin{equation}
\label{sq_2_mode_small_r}
  |{\rm f}\rangle \simeq |0\rangle_1 |0\rangle_2 +\, r\, |1\rangle_1 |1\rangle_2,
\end{equation}
it is easy to show that the state vector $|{\rm f}\rangle$ describes the entangled state. Indeed, the entanglement condition of considering photon state can be verified by investigation of the conditional probability $P\left( {Y_j / X_i } \right) (i,j=\overline{1,2})$ \cite{gluon_entangled,gluon_entangled_NPCS}
\begin{equation}
\label{ent_cond}
P\left( {Y_j / X_i } \right) = \frac{\left\| {\left\langle {Y_j } 
\right|\left\langle {X_i } \right.\left| {{\rm f}} \right\rangle } 
\right\|^2}{\left\langle {{\rm f}} \right|\left. {X_i } \right\rangle 
\left\langle {X_i } \right.\left| {{\rm f}} \right\rangle } = 
\left\{ 
{{\begin{array}{*{20}c}
 {\delta_{ij}} \hfill \\
 \mathrm{or} \hfill \\
 {1 - \delta_{ij}},
\end{array} }} \right.
\end{equation}
where $|X_1\rangle=|0\rangle_1, |X_2\rangle=|1\rangle_1$, $|Y_1\rangle=|0\rangle_2, |Y_2\rangle=|1\rangle_2$.

The dimensionless coefficient
\begin{equation}\label{entangle_coefficient}
y=\left[\frac{|\overline{a_1 a_2^+}|^2 + |\overline{a_1 a_2}|^2}{2(\overline{a_1^+ a_1}+1/2)(\overline{a_2^+ a_2}+1/2)}\right]^{1/2}
\end{equation}
is the measure of entanglement for two-mode states \cite{Dodonov}, $0\leq y <1$ (entanglement is not observe when $y=0$). Here $\overline{a_i a_j^+} = \langle a_i a_j^+\rangle - \langle a_i\rangle\langle a_j^+\rangle$, $a_i, a_j^+$ are the annihilation and creation operators correspondingly. Averaging the annihilation and creation operators in the expression (\ref{entangle_coefficient}) over the vector $|{\rm f}\rangle$ (\ref{sq_2_mode_small_r}) at small squeeze factor we have
\begin{equation}\label{entangle_coefficient_small_squeezing}
y\simeq\sqrt{2} r.
\end{equation}
It is obvious that maximal coefficient of the entanglement is defined by maximal possible value of the squeeze factor $r$\footnote{Here calculations is restricted by first order on the squeeze factor.}: $y_{max}\approx 0.462$ at $r=0.327$.

Two-mode gluon states with two different colours can lead to $q\bar{q}$-entangled states. Interaction of the quark entangled states with stochastic vacuum (quantum measurement) has a remarkable property, namely, as soon as some measurement projects one quark onto a state with definite colour, the another quark also immediately obtains opposite colour that leads to coupling of quark-antiquark pair, string tension inside $q\bar{q}$-pare and free propagation of colourless hadrons. Therefore the investigation of the gluon entangled states is an issue of the day.

\section{Two-mode squeezed gluon states}
In order to verify whether the gluon state vector describes the two-mode squeezed state on colours $h$ and $g$, it is necessary to introduce the phase-sensitive Hermitian operators $(X^{h,g}_{\lambda})_1=\big[b^h_{\lambda}\, +\,b^g_{\lambda}\,+\, b^{h+}_{\lambda}\,+\,b^{g+}_{\lambda}\big] /(2\sqrt{2}) $ and
$(X^{h,g}_{\lambda})_2=\big[b^h_{\lambda}\, +\,b^g_{\lambda}\,-\, b^{h+}_{\lambda}\,-\,b^{g+}_{\lambda}\big] /(2i\sqrt{2}) $ by analogy with quantum optics \cite{QO_Walls} and to establish conditions under which the variance of one of them can be less than the variance of a coherent state. Here $b^h_{(\lambda)}, b^g_{\lambda} (b^{h+}_{\lambda}, b^{g+}_{\lambda})$ are the operators annihilating (creating) of gluons of colours $h,g=\overline{1,8}$ and polarization index $\lambda=\overline{1,3}$.

Mathematically, the condition of two-mode squeezing on colours $h,g$ is expressed in the form of the inequality
\begin{eqnarray}
\label{sq_cond} \Bigl\langle N\left({\Delta}(X^{h,g}_{\lambda})_{1\atop
2}\right)^2 \Bigr\rangle <\, 0.
\end{eqnarray}
Here $N$ is the normal-ordering operator such as
\begin{equation}
\label{form_N} 
\begin{array}{l}
\Bigl\langle N\left(\Delta(X^{h,g}_{\lambda})_{1\atop
2}\right)^2 \Bigr\rangle =\displaystyle\pm\frac18
\Biggl\{\Bigl\langle\Bigl(b^h_{\lambda}\Bigr)^2\Bigr\rangle
-\Bigl\langle b^h_{\lambda}\Bigr\rangle^2 + \Bigl\langle\Bigl(b^g_{\lambda}\Bigr)^2\Bigr\rangle
-\Bigl\langle b^g_{\lambda}\Bigr\rangle^2+
\Bigl\langle\Bigl(b^{h+}_{\lambda}\Bigr)^2\Bigr\rangle
-\Bigl\langle b^{h+}_{\lambda}\Bigr\rangle^2 \\+ \Bigl\langle\Bigl(b^{g+}_{\lambda}\Bigr)^2\Bigr\rangle
-\Bigl\langle b^{g+}_{\lambda}\Bigr\rangle^2 + 2\biggl[\Bigl\langle b^{h}_{\lambda} b^g_{\lambda}\Bigr\rangle\, - \Bigl\langle b^{h}_{\lambda}\Bigr\rangle \Bigl\langle b^g_{\lambda}\Bigr\rangle+\Bigl\langle b^{h+}_{\lambda} b^{g+}_{\lambda}\Bigr\rangle\, - \Bigl\langle b^{g+}_{\lambda}\Bigr\rangle \Bigl\langle b^{g+}_{\lambda}\Bigr\rangle\biggr]\\\pm
2\biggl[\Bigl\langle b^{h+}_{\lambda} b^h_{\lambda}\Bigr\rangle\, - \Bigl\langle b^{h+}_{\lambda}\Bigr\rangle \Bigl\langle b^h_{\lambda}\Bigr\rangle+
\Bigl\langle b^{g+}_{\lambda} b^g_{\lambda}\Bigr\rangle\, - \Bigl\langle b^{g+}_{\lambda}\Bigr\rangle \Bigl\langle b^g_{\lambda}\Bigr\rangle
+\Bigl\langle b^{h+}_{\lambda} b^g_{\lambda}\Bigr\rangle\, - \Bigl\langle b^{h+}_{\lambda}\Bigr\rangle \Bigl\langle b^g_{\lambda}\Bigr\rangle\\\qquad{}+
\Bigl\langle b^{g+}_{\lambda} b^h_{\lambda}\Bigr\rangle\, - \Bigl\langle b^{g+}_{\lambda}\Bigr\rangle \Bigl\langle b^h_{\lambda}\Bigr\rangle\biggr]\!
\Biggr\}.
\end{array}
\end{equation}
The expectation values of the creation and annihilation operators for gluons with specified colour and polarization index are taken over the vector $|{\rm f}\rangle$ which describes the evolution of the virtual gluon field during a small time interval $\Delta t$ within interaction representation
\begin{equation}\label{1H}
|{\rm f}\rangle \simeq |{\rm in}\rangle - \:i\, \Delta t\, H_{\rm I}(t_0) \,|{\rm in}\rangle,
\end{equation}
where $H_{\rm I}(t_0)=H_{\rm I}^{(3)}(t_0)+H_{\rm I}^{(4)}(t_0)$ is the Hamiltonian three-gluon ($H_{\rm I}^{(3)}$) and four-gluon ($H_{\rm I}^{(4)}$) self-interactions the explicit form of which are given in Appendix in momentum representation, $|{\rm in}\rangle$ is an initial state vector of the virtual gluon field, $\Delta t=t-t_0$ (below we will assume $t_0=0$ and consequently $\Delta t=t$).

We choose a product of the coherent states of the gluons with different colour and polarization indexes as the initial state vector, that is $|{\rm in}\rangle \equiv |\,\alpha\rangle= \prod\limits_{\lambda=1}^3\prod\limits_{b=1}^8|\alpha_{\lambda}^b\rangle$ because any vector may be decompose on these basis vectors and coherent states are widely used in QO \cite{QO_Walls,QO_Cambr}. By analogy with QO gluon coherent state vector $|\alpha_{\lambda}^b\rangle$ is the eigenvector of the corresponding annihilation operator $b^b_{\lambda}$ with the eigenvalue $\alpha^b_{\lambda_1}$ which can be written in terms of the gluon coherent field amplitude $|\alpha^b_{\lambda_1}|$ and phase $\gamma_{\lambda}^b$ of the given gluon field  $\alpha^b_{\lambda_1}=|\alpha^b_{\lambda_1}|e^{i\,\gamma^b_{\lambda_1}}$.
In each gluon coherent state $|\alpha_{\lambda}^b\rangle$ the gluon number with fixed colour b and polarization $\lambda$ is arbitrary (the average multiplicity of given gluon is equal to square of the gluon coherent field amplitude $\langle n_{\lambda}^b\rangle= |\alpha_{\lambda}^b|^2$) and phase of considering state $\gamma_{\lambda}^b$ is fixed.

Averaging the annihilation and creation operators $b^h_{\lambda}, b^g_{\lambda}, b^{h+}_{\lambda}, b^{g+}_{\lambda}$ in (\ref{form_N}) over the evolved vector $|{\rm f}\rangle$ which is defined according to (\ref{1H}) and taking into account of chosen initial state vector we write the two-mode squeezing condition in the form
\begin{eqnarray}
\label{sq_cond1}
\Bigl\langle N\left(\Delta(X^{h,g}_{\lambda})_{1\atop
2}\right)^2 \Bigr\rangle & =&\pm\frac{i\, t}{8}\biggl\{\bigl\langle\alpha\bigl|\bigl[\bigl[H_{\rm I}(0), b^{h+}_{\lambda}\bigr], b^{h+}_{\lambda}\bigr]\bigr|\alpha\bigr\rangle
-\bigl\langle\alpha\bigl|\bigl[b^{h}_{\lambda}, \bigl[b^{h}_{\lambda}, H_{\rm I}(0)\bigr]\bigr]\bigr|\alpha\bigr\rangle\nonumber \\&&{}+ \bigl\langle\alpha\bigl|\bigl[\bigl[H_{\rm I}(0), b^{g+}_{\lambda}\bigr], b^{g+}_{\lambda}\bigr]\bigr|\alpha\bigr\rangle
-\bigl\langle\alpha\bigl|\bigl[b^{g}_{\lambda}, \bigl[b^{g}_{\lambda}, H_{\rm I}(0)\bigr]\bigr]\bigr|\alpha\bigr\rangle\nonumber \\&&{}+2\bigl
\langle\alpha\bigl|\bigl[\bigl[H_{\rm I}(0), b^{h+}_{\lambda}\bigr], b^{g+}_{\lambda}\bigr]\bigr|\alpha\bigr\rangle
-2\bigl\langle\alpha\bigl|\bigl[b^{g}_{\lambda}, \bigl[b^{h}_{\lambda}, H_{\rm I}(0)\bigr]\bigr]\bigr|\alpha\bigr\rangle\biggr\} < 0.
\end{eqnarray}
It is easy to show that the three-gluon self-interaction (as for single-mode squeezing of the gluons \cite{shaporov}) does not lead to squeezing effect since
\begin{equation}
\label{sq_cond_three_gluon}
\left.
\begin{array}{l}
\bigl[\bigl[H_{\rm I}^{(3)}(0), b^{h+}_{\lambda}\bigr], b^{h+}_{\lambda}\bigr]=0,\quad \bigl[\bigl[H_{\rm I}^{(3)}(0), b^{g+}_{\lambda}\bigr], b^{g+}_{\lambda}\bigr]=0,\quad
\bigl[\bigl[H_{\rm I}^{(3)}(0), b^{h+}_{\lambda}\bigr],b^{g+}_{\lambda}\bigr]=0,\\\bigl[b^{g}_{\lambda}, \bigl[b^{h}_{\lambda}, H_{\rm I}^{(3)}(0)\bigr]\bigr]=0,\quad\quad \bigl[b^{h}_{\lambda}, \bigl[b^{h}_{\lambda}, H_{\rm I}^{(3)}(0)\bigr]\bigr]=0,\quad\quad \bigl[b^{g}_{\lambda}, \bigl[b^{g}_{\lambda}, H_{\rm I}^{(3)}(0)\bigr]\bigr]=0.
\end{array}
\right\}
\end{equation}
Thus only the four-gluon self-interaction can yield a two-mode squeezing effect. Indeed, the two-mode squeezing condition can be written in the explicit form as
\begin{equation}
\label{sq_cond_four_gluon}
\begin{array}{l}
\Bigl\langle N\left(\Delta(X^{h,g}_{\lambda})_{1\atop
2}\right)^2 \Bigr\rangle =\displaystyle\pm \frac{i\, t}{8} g^2 (2\pi)^3 \int d\tilde{k}_1 d\tilde{k}_2 \sum\limits_{\lambda_1, \lambda_2}
\times\Biggl\{\langle \alpha |b_{\lambda_1}^{b+}(k_1)b_{\lambda_2}^{c+}(k_2)-b_{\lambda_1}^{b}(k_1)
b_{\lambda_2}^{c}(k_2)|\alpha\rangle\\ \times\Bigl[\delta(2\vec{k}-\vec{k_1}-\vec{k_2})-\delta(2\vec{k}+\vec{k_1}
+\vec{k_2})\Bigr]\, \Bigl[(f_{ahb}f_{ahc}
+f_{agb}f_{agc}+2f_{ahb}f_{agc})
\Bigl(\varepsilon_\mu^{\lambda_1}(k_1)\,
\varepsilon_{\lambda_2}^\mu(k_2)\\\times
\varepsilon_\nu^{\lambda}(k)\,
\varepsilon_{\lambda}^\nu(k)
-\varepsilon_\mu^{\lambda_1}(k_1)
\varepsilon_{\lambda}^\mu(k)\varepsilon_\nu^{\lambda_2}(k_2)\,
\varepsilon_{\lambda}^\nu(k)
\Bigr) 
 -\, 2f_{ahg}f_{abc}\varepsilon_\mu^{\lambda_1}(k_1)
\varepsilon_{\lambda}^\mu(k)\varepsilon_\nu^{\lambda_2}(k_2)
\varepsilon^\nu_{\lambda}(k)\Bigr]
\\+\,2\langle \alpha |b_{\lambda_1}^{b+}(k_1)b_{\lambda_2}^{c}(k_2)|\alpha\rangle
\times\Bigl[\delta(2\vec{k}-\vec{k_1}+\vec{k_2})-\delta(2\vec{k}+\vec{k_1}
-\vec{k_2})\Bigr](f_{ahb}f_{ahc}
+f_{agb}f_{agc}\\+f_{ahc}f_{agb}+f_{ahb}f_{agc})\Bigl(
\varepsilon_\mu^{\lambda_1}(k_1)\,
\varepsilon_{\lambda_2}^\mu(k_2)\varepsilon_\nu^{\lambda}(k)\,
\varepsilon_{\lambda}^\nu(k)
-\varepsilon_\mu^{\lambda_1}(k_1)\,
\varepsilon_{\lambda}^\mu(k)\varepsilon_\nu^{\lambda_2}(k_2)\,
\varepsilon_{\lambda}^\nu(k)
\Bigr)
\Biggr\}<0.
\end{array}
\end{equation}
Here $g$ is a self-interaction constant, $d\tilde{k} = \displaystyle\frac{d^3 k }{(2\pi)^3 2k_0}$, $k_0$ is a gluon energy, $\varepsilon^\mu_\lambda$ is a polarization vector, $f_{ahb}$ are structure constants of $\rm SU_c(3)$ group.

For visualization, let us investigate the obtained two-mode squeezing condition (\ref{sq_cond_four_gluon}) for collinear gluon. In this case corresponding squeezing condition is
\begin{eqnarray}
\label{sq_cond_collin_gluon}
\Bigl\langle N\left(\Delta(X^{h,g}_{\lambda})_{1\atop
2}\right)^2 \Bigr\rangle &=&\,\displaystyle\pm\, t\,\frac{\alpha_s\pi}{4 k_0} (f_{ahb}f_{ahc}+f_{agb}f_{agc}+f_{ahb}f_{agc}+
f_{agb}f_{ahc})\nonumber\\&&{}\times
\sum\limits_{\lambda_1\neq\lambda}
|\alpha^b_{\lambda_1}||\alpha^c_{\lambda_1}|
\sin(\gamma^b_{\lambda_1}+\gamma^c_{\lambda_1}) < 0.
\end{eqnarray}

\vspace{0.2cm}
\noindent
Here we have taken into account that $\alpha^b_{\lambda_1}=|\alpha^b_{\lambda_1}|e^{i\,\gamma^b_{\lambda_1}}$ and $\alpha^c_{\lambda_1}=|\alpha^c_{\lambda_1}|e^{i\,\gamma^c_{\lambda_1}}$, $\alpha_s = g^2/(4\pi)$ is a coupling constant\footnote{Here we also take into consideration the gluons with momenta directed toward jet development.}. The two-mode squeezing condition is fulfilled for any cases apart from $\gamma^b_{\lambda_1}+\gamma^c_{\lambda_1} = 0, \pi$. In particular, if all initial gluon coherent fields are real or imaginary then the two-mode squeezing condition is not fulfilled as in the single-mode case. Obviously, the larger are both amplitudes of the initial gluon coherent fields with different colour and polarization indexes and coupling constant, the larger is two-mode squeezing effect.

Thus non-perturbative gluon evolution (at large coupling constant) is very significant under investigation of the squeezing effect.

\section{Entangled collinear gluon states}
By analogy with QO we assume that two-mode gluon SS with fixed colours $h,g$ is closely connected with corresponding entangled states of the gluons. 

At small value of the squeeze factor we have
\begin{equation}\label{r}
  r =2\,\left|\Bigl\langle N\left(\Delta(X^{h,g}_{\lambda})_{1\atop
2}\right)^2 \Bigr\rangle\right|.
\end{equation}
As initial states we can take a vector including $\mid\!0_{\lambda}^h\rangle\mid\!0_{\lambda}^g\rangle$ evolution of which at small time can be written in terms of the squeeze factor as
\begin{equation}
\label{sq_gluon_small_r}
  |{\rm f}\rangle \simeq |0_\lambda^h\rangle |0_\lambda^g\rangle +\, r\, |1_\lambda^h\rangle |1_\lambda^g\rangle,
\end{equation}
where the squeeze factor for the collinear gluons is defined according to (\ref{r}) and (\ref{sq_cond_collin_gluon}).

The entanglement condition of considering gluon states with colours $h,g$ and polarization $\lambda$ can be verified by investigation of the conditional probability $P\left( {Y_j / X_i } \right) (i,j=\overline{1,2})$ by analogy with corresponding condition for photons (\ref{ent_cond}) assuming that
$|X_1\rangle~=~|0_\lambda^h\rangle$, $|X_2\rangle~=~|1_\lambda^h\rangle$, $|Y_1\rangle~=~|0_\lambda^g\rangle,~|Y_2\rangle~=~|1_\lambda^g\rangle$.
It is not complicated to make sure that the condition (\ref{ent_cond}) is fulfilled for the cases as for the two-mode squeezing for the state vector $|{\rm f}\rangle$ (\ref{sq_gluon_small_r}) which describes the non-perturbative evolution of the gluon fields during small time.

Rewriting the expression (\ref{entangle_coefficient}) for the entanglement coefficient in case two-mode gluon states as
\begin{equation}\label{entanglement_coeff_gluon}
y=\left[\frac{|\overline{b_\lambda^h b_\lambda^{g+}}|^2 + |\overline{b_\lambda^h b_\lambda^g}|^2}{2(\overline{b_\lambda^{h+} b_\lambda^h}+1/2)(\overline{b_\lambda^{g+} b_\lambda^g}+1/2)}\right]^{1/2}
\end{equation}
we can write the condition of the entanglement with taking into account (\ref{r}) and (\ref{sq_cond_collin_gluon})
\begin{equation}\label{condition_max_entanglement_gluon}
0< \left| t\,\frac{\alpha_s\pi}{\sqrt{2} k_0} (f_{ahb}f_{ahc}+f_{agb}f_{agc}+f_{ahb}f_{agc}+
f_{agb}f_{ahc})
\sum\limits_{\lambda_1\neq\lambda}
|\alpha^b_{\lambda_1}||\alpha^c_{\lambda_1}|
\sin(\gamma^b_{\lambda_1}+\gamma^c_{\lambda_1})\right| < 1.
\end{equation}

Thus by analogy with quantum optics as a result of four-gluon self-interaction we obtain two-mode squeezed gluon states which are also entangled.

\section{Conclusion}
Investigating of the gluon fluctuations we have proved theoretically the possibility of existence of the gluon two-mode squeezed states. The emergence of such remarkable states becomes possible owing to the four-gluon self-interaction. The three-gluon self-interaction does not lead to squeezing effect.

We have shown that QCD jet non-perturbative evolution leads both to squeezing and entanglement of gluons. It should be noted that the greater are both the amplitudes of the initial gluon coherent fields with different colour and polarization indexes and coupling constant, the greater is squeezing effect of the colour gluons. We have demonstrated that entanglement condition of the  gluon states with fixed two colours and polarization is defined by the corresponding squeeze factor which is dependent on the amplitudes and phases of initial coherent gluon fields.

Two-mode gluon states with two different colours can lead to $q\bar{q}$-entangled states role of which could be very significant for explanation of the confinement phenomenon.

\section*{Appendix}
\[
\begin{array}{l}
H_{\rm I}^{(3)}(t) = \displaystyle ig(2\pi)^3 f_{abc}\sum\limits_{\lambda_1,\lambda_2,\lambda_3}\int d\tilde{k}_1 d\tilde{k}_2 d\tilde{k}_3
\times\,\biggl\{
\vec{k}_1\vec{\varepsilon}_{\lambda_2}(k_2)
\varepsilon_{\lambda_1}^\nu(k_1)\varepsilon_\nu^{\lambda_3}(k_3)
\delta(\vec{k}_1+\vec{k}_2+\vec{k}_3)
\\\vspace{-0.3cm}\\
\qquad\times\,\bigl[b^a_{\lambda_1}(k_1)b^b_{\lambda_2}(k_2)b^c_{\lambda_3}(k_3)\;
e^{-2i(k_{01}+k_{02}+k_{03})t}
-\,b^{a+}_{\lambda_1}(k_1)
b^{b+}_{\lambda_2}(k_2)b^{c+}_{\lambda_3}(k_3)\:
e^{2i(k_{01}+k_{02}+k_{03})t}
\bigr]
\\\vspace{-0.3cm}\\\qquad+\,b^{a+}_{\lambda_1}(k_1)b^{b}_{\lambda_2}(k_2)
b^{c}_{\lambda_3}(k_3)\:
e^{2i(k_{01}-k_{02}-k_{03})t}\;\delta(\vec{k}_1-\vec{k}_2-\vec{k}_3)
\\\vspace{-0.3cm}\\\qquad\times
\Bigl[\vec{k}_3
\vec{\varepsilon}_{\lambda_1}(k_1)
\varepsilon_{\lambda_2}^\nu(k_2)\varepsilon_\nu^{\lambda_3}(k_3)
+\vec{k}_2
\vec{\varepsilon}_{\lambda_3}(k_3)
\varepsilon_{\lambda_1}^\nu(k_1)\varepsilon_\nu^{\lambda_2}(k_2)
-\vec{k}_1
\vec{\varepsilon}_{\lambda_2}(k_2)
\varepsilon_{\lambda_1}^\nu(k_1)\varepsilon_\nu^{\lambda_3}(k_3)\Bigr]
\\\vspace{-0.3cm}\\\qquad
+\,b^{a+}_{\lambda_1}(k_1)b^{b+}_{\lambda_2}(k_2)b^{c}_{\lambda_3}(k_3)
\,e^{2i(k_{01}+k_{02}-k_{03})t}\;\delta(\vec{k}_1+\vec{k}_2-\vec{k}_3)
\\\vspace{-0.3cm}\\\qquad\times
\Bigl[\vec{k}_1
\vec{\varepsilon}_{\lambda_3}(k_3)
\varepsilon_{\lambda_1}^\nu(k_1)\varepsilon_\nu^{\lambda_2}(k_2)
+\vec{k}_3
\vec{\varepsilon}_{\lambda_1}(k_1)
\varepsilon_{\lambda_2}^\nu(k_2)\varepsilon_\nu^{\lambda_3}(k_3)
-\vec{k}_1
\vec{\varepsilon}_{\lambda_2}(k_2)
\varepsilon_{\lambda_1}^\nu(k_1)\varepsilon_\nu^{\lambda_3}(k_3)\Bigr]
\biggr\},\hspace{0.1cm} (A.1)
\end{array}
\]
\[
\begin{array}{l}
H_{\rm I}^{(4)}(t) = \displaystyle \frac{g^2}4(2\pi)^3 \sum\limits_{\lambda_1,\lambda_2,\lambda_3,\lambda_4}\int d\tilde{k}_1 d\tilde{k}_2 d\tilde{k}_3 d\tilde{k}_4\biggl\{
\varepsilon_{\lambda_1}^\mu(k_1)\varepsilon_\mu^{\lambda_3}(k_3)
\varepsilon_{\lambda_2}^\nu(k_2)\varepsilon_\nu^{\lambda_4}(k_4) f_{abc}f_{ade}\\\vspace{-0.3cm}\\\qquad\times
\Bigl[\delta(\vec{k}_1+\vec{k}_2+\vec{k}_3+\vec{k}_4)
\Bigl(b^b_{\lambda_1}(k_1)b^c_{\lambda_2}(k_2)b^d_{\lambda_3}(k_3)
b^e_{\lambda_4}(k_4)\: e^{-2i(k_{01}+k_{02}+k_{03}+k_{04})t}\\\vspace{-0.3cm}\\\qquad+\,
b^{b+}_{\lambda_1}(k_1)b^{c+}_{\lambda_2}(k_2)b^{d+}_{\lambda_3}(k_3)
b^{e+}_{\lambda_4}(k_4)\:e^{2i(k_{01}+k_{02}+k_{03}+k_{04})t}\Bigr)
\\\vspace{-0.3cm}\\\qquad+\,4b^{b+}_{\lambda_1}(k_1)
b^c_{\lambda_2}(k_2)b^d_{\lambda_3}(k_3)
b^e_{\lambda_4}(k_4)\:e^{2i(k_{01}-k_{02}-k_{03}-k_{04})t}\,
\delta(\vec{k}_1-\vec{k}_2-\vec{k}_3-\vec{k}_4)\\\vspace{-0.3cm}\\\qquad
+\, 4b^{b+}_{\lambda_1}(k_1)b^{c+}_{\lambda_2}(k_2)b^{d+}_{\lambda_3}(k_3)
b^e_{\lambda_4}(k_4)\:e^{2i(k_{01}+k_{02}+k_{03}-k_{04})t}
\,\delta(\vec{k}_1+\vec{k}_2+\vec{k}_3-\vec{k}_4)\Bigr]\\\vspace{-0.3cm}\\
\qquad+\,2b^{b+}_{\lambda_1}(k_1)b^{c+}_{\lambda_2}(k_2)b^d_{\lambda_3}(k_3)
b^e_{\lambda_4}(k_4)\:e^{2i(k_{01}+k_{02}-k_{03}-k_{04})t}
\,\delta(\vec{k}_1+\vec{k}_2-\vec{k}_3-\vec{k}_4)\\\vspace{-0.3cm}\\
\qquad\times \Bigl[\varepsilon_{\lambda_1}^\mu(k_1)\varepsilon_\mu^{\lambda_3}(k_3)
\varepsilon_{\lambda_2}^\nu(k_2)\varepsilon_\nu^{\lambda_4}(k_4)
(f_{abc}f_{ade}+f_{abe}f_{adc})\\\vspace{-0.3cm}\\\qquad\quad{}+
\varepsilon_{\lambda_1}^\mu(k_1)\varepsilon_\mu^{\lambda_2}(k_2)
\varepsilon_{\lambda_3}^\nu(k_3)\varepsilon_\nu^{\lambda_4}(k_4)f_{abd}f_{ace}
\Bigr]
\biggr\}.\hspace{6.7cm} (A.2)
\end{array}
\]

\end{document}